\documentclass[pdflatex,sn-basic]{sn-jnl}

\usepackage{graphicx}%
\usepackage{multirow}%
\usepackage{amsmath,amssymb,amsfonts}%
\usepackage{amsthm,amsfonts,amsmath,amssymb,bm}%
\usepackage{mathrsfs}%
\usepackage[title]{appendix}%
\usepackage{xcolor}%
\usepackage{textcomp}%
\usepackage{manyfoot}%
\usepackage{booktabs}%
\usepackage{algorithm}%
\usepackage{algorithmicx}%
\usepackage{algpseudocode}%
\usepackage{listings}%
\usepackage{ulem}%

\usepackage{multirow} 
\usepackage{comment}
\usepackage{hyperref}
\usepackage{graphicx}
\usepackage{float}

\def\apj{{Astroph.\@ J.\ }}

\def\mnras{{Mon.\@ Not.\@ Roy.\@ Ast.\@ Soc. }}

\def\prd{{Phys.\@ Rev.\@ D\ }}

\def\apjl{{Astroph.\@ J.\@ Lett.}}
\def\jcap{{JCAP}}

\def\apjs{ApJS}
\def\aap{AAp}
\def\physscr{PhyS}
\def\aas{AJ}

\renewcommand{\textbf}[1]{#1}

\makeatletter
\renewcommand{\section}{\@startsection{section}{1}{\z@}%
	{-3.5ex \@plus -1ex \@minus -.2ex}%
	{2.3ex \@plus.2ex}%
	{\normalfont\Large\bfseries}}%

\renewcommand{\subsection}{\@startsection{subsection}{2}{\z@}%
	{-3.25ex\@plus -1ex \@minus -.2ex}%
	{1.5ex \@plus .2ex}%
	{\normalfont\large\bfseries}}%

\renewcommand{\subsubsection}{\@startsection{subsubsection}{3}{\z@}%
	{-3.25ex\@plus -1ex \@minus -.2ex}%
	{1.5ex \@plus .2ex}%
	{\normalfont\normalsize\bfseries}}%
\makeatother

\theoremstyle{thmstyleone}%
\theoremstyle{thmstyletwo}%

\theoremstyle{thmstylethree}%

\newcommand{\secref}[1]{\hyperref[#1]{Section~\ref*{#1}}}
\newcommand{\appref}[1]{\hyperref[#1]{Appendix~\ref*{#1}}}

\begin{document}

\title[Low Redshift Observational Constraints on Dark Energy Models using ANN - CosmicANNEstimator]{Low Redshift Observational Constraints on Dark Energy Models using ANN - CosmicANNEstimator}

\author[1]{\fnm{Ashly} \sur{Joseph}}

\author*[2]{\fnm{Albin} \sur{Joseph}}\email{ajosep52@asu.edu}

\author[1]{\fnm{Christina} \sur{Terese Joseph}}

\author[1]{\fnm{John} \sur{Paul Martin}}

\author[3]{\fnm{Sunil} \sur{Kumar PV}}

\author[1]{\fnm{Sarthak} \sur{Giri}}

\affil[1]{\orgdiv{Department of Computer Science, IIITKottayam, Kerala, 686635, India}}

\affil[2]{\orgdiv{Department of Physics, Arizona State University, Tempe, AZ 85287, USA}}

\affil[3]{\orgdiv{Department of Computer Science, IIITDharwad, Karnataka, 580009, India}}

\abstract{We present CosmicANNEstimator (Cosmological Parameters Artificial Neural Network Estimator), a  machine learning approach for constraining cosmological parameters within the Lambda Cold Dark Matter ($\Lambda$CDM) framework. Our methodology employs two specialized artificial neural networks (ANNs) designed to analyze Hubble parameter and Supernova data independently. The estimator is trained on synthetic data covering broad parameter ranges, with Gaussian random noise incorporated to simulate observational uncertainties. Our results demonstrate parameter estimates and associated uncertainties comparable to traditional Markov Chain Monte Carlo (MCMC) methods, establishing machine learning as an efficient  alternative for cosmological parameter estimation. This work underscores the potential of neural network–based inference to complement traditional Bayesian methods and accelerate future cosmological analyses.
}


\maketitle

\section{Introduction}\label{sec1}

Modern cosmology has witnessed unprecedented advances over the past two decades, driven by a remarkable wealth of high-precision observational data. This wealth of information has enabled rigorous testing and refinement of cosmological models, deepening our understanding of universal evolution. The Lambda Cold Dark Matter ($\Lambda$CDM) model has emerged as the predominant theoretical framework, emphasizing two fundamental components: cold dark matter, characterized by its slow-moving nature, and dark energy, a mysterious force that drives the accelerating expansion of the universe while opposing gravitational forces ~\citep{1999ApJ...517..565P,2007ApJS..170..377S,2004PhRvD..69j3501T,1998JETP...87..223Z,2007ApJ...666..716D,2009ApJ...700.1097H,2009ApJS..180..225H,2000MNRAS.317..893L,2010ApJ...714L.185B,2021JApA...42..111J,2022MNRAS.511.1637J,2023MNRAS.519.1809J}.

The quantification  of cosmological parameters serves as a detailed blueprint of universal properties, providing crucial insights into cosmic evolution across temporal scales. These parameters form the essential bridge between theoretical frameworks and observational evidence, making their accurate estimation a fundamental pursuit in modern cosmology. The Markov Chain Monte Carlo (MCMC) method has traditionally been the cornerstone of parameter estimation, valued for its robust performance in both parameter value determination and uncertainty quantification.
However, MCMC methodology faces significant challenges in the era of big data. As observational datasets grow in both volume and complexity, the computational demands of MCMC become increasingly prohibitive. The method's inherent requirement for numerous iterations to achieve proper convergence and thoroughly explore parameter space poses practical limitations for modern cosmological analyses. This computational bottleneck has spurred the search for more efficient analytical approaches.
In response to these challenges, this work investigates an artificial intelligence-based solution for cosmological parameter estimation. While the use of ANNs for cosmological estimations is not novel, this work focuses on applying the ANN methodology to multiple datasets—specifically, Hubble data and supernova data.

Machine learning (ML) methods have revolutionized cosmological research in recent years, demonstrating exceptional performance in addressing complex challenges with both accuracy and efficiency~\citep{2022Univ....8..120D}. The application of ML regression algorithms—including Artificial Neural Networks, KNN, Gradient Boosting, Extra-Trees and Support Vector Machines—has proven particularly effective in estimating \(H_0\)  and its associated uncertainties ~\citep{2023EPJC...83..548B}. Genetic Algorithms have emerged as a valuable complement to traditional MCMC methods, especially in constraining dark energy models such as \(\Lambda\)CDM, CPL, and PolyCPL~\citep{2023Univ...10...11M}.

Significant methodological advances have shaped the field's development. Auld et~al.\ introduced COSMONET~\citep{2007MNRAS.376L..11A}, a  Bayesian inference algorithm that employs artificial neural networks (ANNs) to accelerate the calculation of Cosmic Microwave Background (CMB)  likelihood functions and matter power spectra for estimating of cosmological parameter. Additionally, Graff et~al.\ developed the blind accelerated multimodal Bayesian inference (BAMBI) algorithm~\citep{2014MNRAS.441.1741G}, which  combines nested sampling with ANNs for efficient likelihood function learning.

In the realm of Hubble parameter analysis, diverse analytical approaches have yielded valuable insights. Leaf and Melia  ~\citep{2017MNRAS.470.2320L} employed two-point statistics to evaluate cosmological models, focusing particularly on differential age (DA) measurements of H(z). Their work proved especially significant due to the model-independent nature of cosmic chronometer measurements. Extending this research, Geng em et~al.\ ~\citep{2018CoTPh..70..445G} combined DA and BAO (baryon acoustic oscillation) techniques to study H(z) measurements, both for parameter determination and to assess the anticipated impact of next-generation  H(z) measurements on parameter estimation.
Latest literature has increasingly focused on ML applications in low-redshift data analysis. Notable contributions by Arjona \& Nesseris~\citep{ 2020PhRvD.101l3525A}, Mukherjee em et~al.\ ~\citep{ 2022JCAP...12..029M}, and Gómez-Valent \& Amendola ~\citep{2018JCAP...04..051G} have successfully employed ML approaches to analyze Hubble measurements and Type Ia Supernovae observations  to refine current cosmological parameter estimates.

The present work draws primary inspiration from two recent frameworks: EcoPANN ~\citep{2020ApJS..249...25W} and ParamANN \citep{2024PhyS...99k5007P}. EcoPANN provides comprehensive estimation of cosmological parameters including \(H_0\), \(\Omega_{bh^2}\), \(\Omega_{ch^2}\), \(\tau\), \(10^9A_s\), and \(n_s\), utilizing an innovative iterative approach to handle uncertainty in diverse cosmological datasets and predict joint constraints. ParamANN focuses on predicting the Hubble constant(\(H_0\)) and density parameters (\(\Omega_{m0}\), \(\Omega_{k0}\), \(\Omega_{\Lambda}\)) of the \(\Lambda\)CDM model along with their uncertainties (\(\sigma_{H_0}\), \(\sigma_{\Omega_{m0}}\), \(\sigma_{\Omega_{k0}}\), \(\sigma_{\Omega_{\Lambda}}\)), employing Hubble measurements and a heteroscedastic loss function for uncertainty quantification.

This study aims to demonstrate the viability of ANNs with varying architectures as alternatives to traditional MCMC methods for cosmological parameter estimation. We employ heteroscedastic loss function to predict three fundamental cosmological parameters ($H_0$, $\Omega_{m0}$, and  $\Omega_{\Lambda0}$) using Hubble data. Also we extend this framework by developing an independent model trained specifically on Type Ia supernova data to predict two key parameters ($\Omega_{m0}$, and  $\Omega_{\Lambda0}$), utilizing the same heteroscedastic uncertainty quantification approach. This dual-model strategy represents a significant departure from  single-dataset approaches, enabling each model to extract and leverage distinct insights from its respective observational dataset.

\subsection{Cosmological Dynamics}

Cosmic evolutionary mechanisms are well understood within the framework of general
relativity (GR)~\citep{2003astro.ph..1448K}. The metric evolves according to the Einstein field equations,
which relate the geometry of the spacetime with the matter content within it.
The Einstein field equations which govern the dynamics of
the Universe are given by~\citep{2003astro.ph..1448K},
\begin{equation}
	G_{\mu\nu} \equiv R_{\mu\nu} - \frac{1}{2} g_{\mu\nu} R = 8 \pi G T_{\mu\nu}, 
	\label{eq:einstein_field}
\end{equation}
where \( G_{\mu\nu} \) is the Einstein tensor,\( G \) is the Newtonian gravitational constant, \( R_{\mu\nu} \) is the Ricci tensor and R is the Ricci scalar . The geometry of spacetime is described by the
left-hand side of the equation, while the energy-momentum content of the Universe
is represented by the right-hand side through the energy-momentum tensor \( T_{\mu\nu} \).

Using Eqn.~\ref{eq:einstein_field}, we can obtain Friedmann equations which govern dynamics of the universe given by,
\begin{equation}
	\left( \frac{\dot{a}}{a} \right)^2 = \frac{8\pi G}{3} \rho - \frac{k}{a^2},
	\label{eq:friedmann1}
\end{equation}
\begin{equation}
	\frac{\ddot{a}}{a} = -\frac{4\pi G}{3} (\rho + 3P).
	\label{eq:friedmann2}
\end{equation}
In this context, ~\(\rho\) and \(P\)  represent the aggregate energy density and pressure components in the universe. We denote  \(\rho_m\) for the contribution by matter (with \(\rho_c\) for cold dark matter and \(\rho_b\)
for baryons), and \(\rho_\Lambda\) signifies  the vacuum energy contribution and \(\rho_r\) represents  the contribution from radiation (with \(\rho_\gamma\) for photons
and \(\rho_\nu\) for neutrinos). The primary Friedmann equation is traditionally expressed via the Hubble parameter, ~\hbox{$H \equiv \dot{a}/a$}
\begin{equation}
	H^2 = \frac{8\pi G}{3} \rho - \frac{k}{a^2}.
	\label{eq:hubble_param}
\end{equation}
In the case of a flat universe (\( k = 0 \)), the critical density of the Universe can be expressed as,

\begin{align}
	\rho_{\text{crit}0} &= \frac{3H_0^2}{8\pi G}\,,
	\label{eq:critical_density}
\end{align}
where the subscript `0` denotes the present-epoch value of cosmological parameters.

The critical density serves as the reference point for expressing density parameters in dimensionless form:
\begin{equation}
	\Omega_{I0} \equiv \frac{\rho_{I0}}{\rho_{\text{crit}0}}.
	\label{eq:density_param}
\end{equation}
From this perspective, the Friedmann equation, (Eqn.~\ref{eq:hubble_param}) takes the following form 

\begin{align}
	H^2(a) &= H_0^2 \bigg[ \Omega_{r0} \left( \frac{a_0}{a} \right)^4 
	+ \Omega_{m0} \left( \frac{a_0}{a} \right)^3 \nonumber \\
	&\quad + \Omega_{k0} \left( \frac{a_0}{a} \right)^2 + \Omega_{\Lambda0} \bigg],
	\label{eq:friedmann_density}
\end{align}

where we have defined a curvature density parameter,~\( \Omega_{k0} \equiv -k/(a_0 H_0)^2 \), $H_0$ is the present-day value of the Hubble parameter (the Hubble constant), $\Omega_{r0}$ denotes the present-day radiation density parameter for matter, $\Omega_{m0}$ denotes the present-day density parameter for matter and $\Omega_{\Lambda0}$ denotes the present-day density parameter for dark energy.
By adopting the usual convention, we normalize the present-day scale factor to unity, \( a_0 \equiv 1 \). Eqn.~\ref{eq:friedmann_density} then becomes
\begin{equation}
	\frac{H^2}{H_0^2} = \Omega_r a^{-4} + \Omega_m a^{-3} + \Omega_k a^{-2} + \Omega_\Lambda.
	\label{eq:friedmann_final}
\end{equation}

Using Eqn.~\ref{eq:friedmann2}, the Friedmann equation of the $\Lambda$CDM model in the late universe (we can neglect $\Omega_{r0}$) with spatial curvature can be written as:

\begin{align}\label{eqn1}
	H^2(z) = H_0^2 \Bigl[ \Omega_{m0} (1 + z)^3 &+ (1 - \Omega_{m0} - \Omega_{\Lambda 0}) (1 + z)^2 \nonumber \\
	&+ \Omega_{\Lambda 0} \Bigr]
\end{align}

where, $H(z)$ represents the Hubble parameter as a function of redshift. This equation \ref{eqn1} illustrates the mathematical model of Hubble parameters across different redshifts based on specific values of  $H_0$, $\Omega_{m0}$, and $\Omega_{\Lambda0}$.

\section{Methodology}

\begin{figure*}[h]
	\centering
	\includegraphics[width=0.7\linewidth]{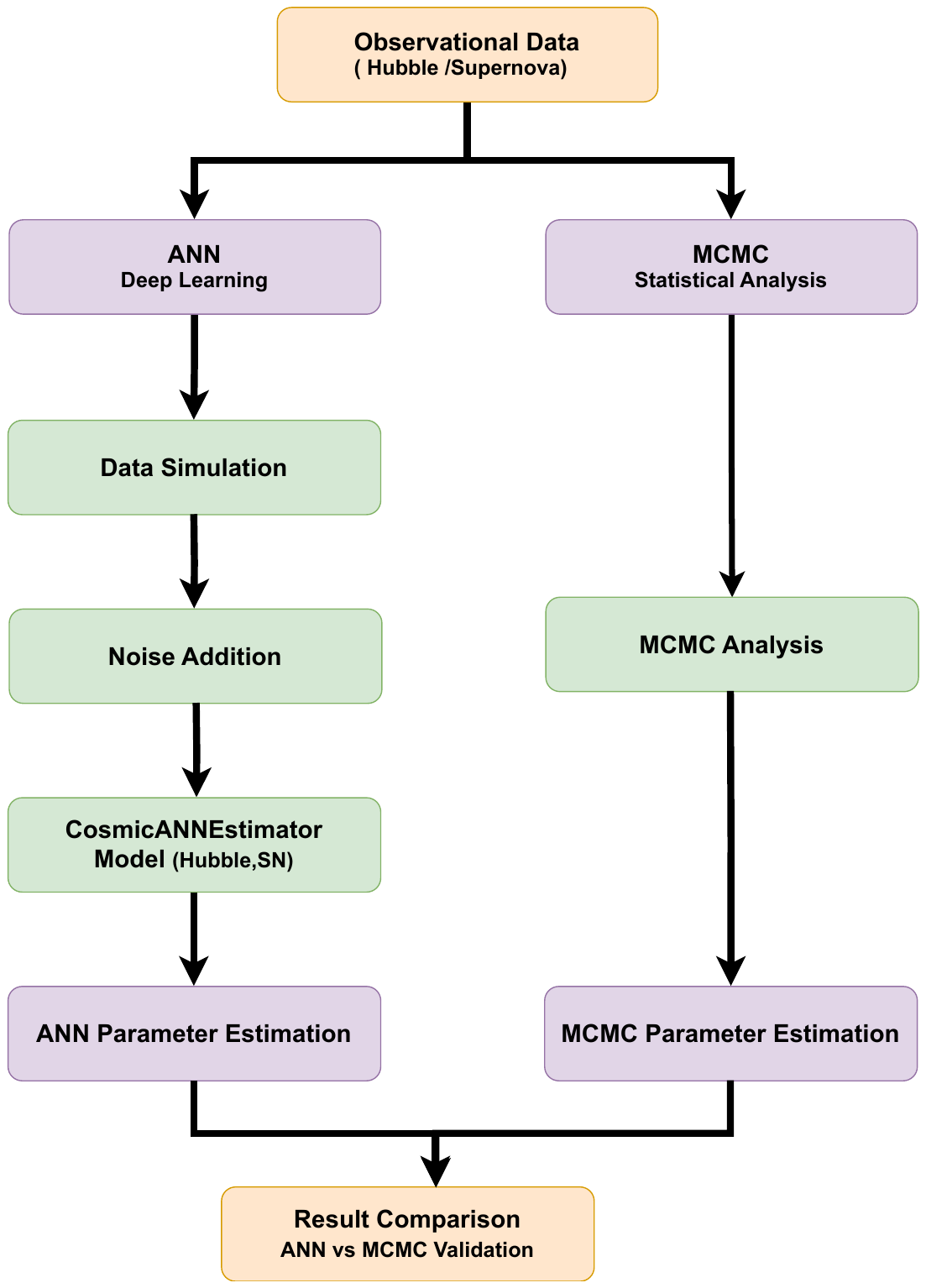}
	\caption{Schematic representation of the CosmicANNEstimator framework integrating observational data processing, neural network training, and MCMC validation}
	\label{fig:METHOD}
\end{figure*}

Our study's methodological approach integrates four key components: First, we examine observational data obtained from Hubble measurements and Pantheon Type Ia Supernovae (SnIa). Second, we generate simulated values for the Hubble parameter  \(H(z)\)  and luminosity distance  \( d_{L}(z) \) , incorporating appropriate noise models. Third, we design and implement specialized neural network architectures for parameter estimation. Finally, we employ Markov Chain Monte Carlo (MCMC) analysis as a validation framework to verify our results.\textbf{ Figure~\ref{fig:METHOD} shows the schematic representation of our approach.}

\subsection{Datasets}
The datasets utilized in this work contain redshifts, an observable associated with each redshift and their respective statistical errors. Each redshift data point is accompanied by an associated observable—such as luminosity distance \( d_L(z) \) or Hubble parameter \( H(z) \)—and their respective statistical errors, which quantify the uncertainties in these measurements.

\begin{figure*}[htp]  
	\centering
	\hspace*{-1.5cm}\includegraphics[width=1.2 \textwidth]{"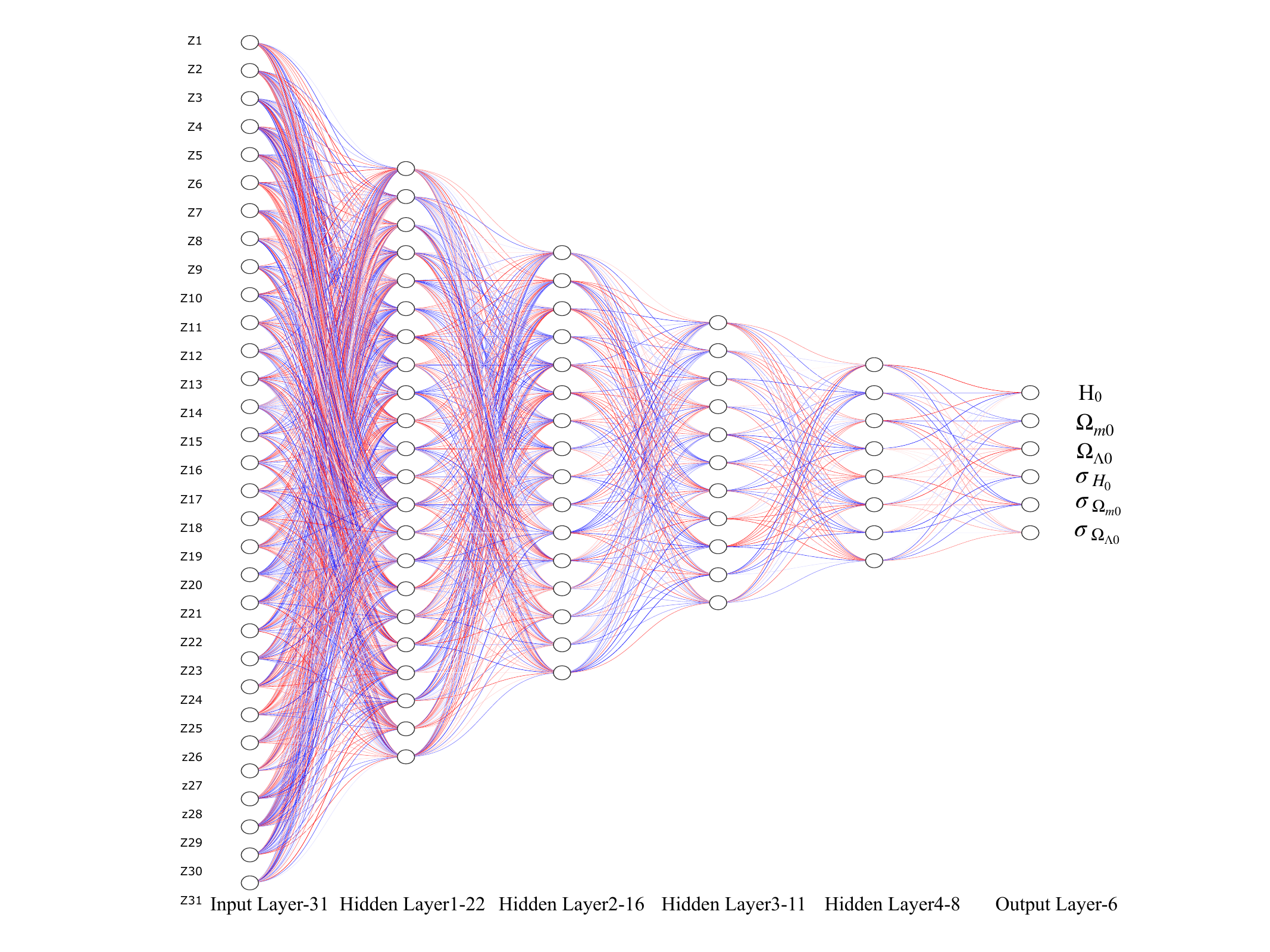"}
	\caption{The architecture of CosmicANNEstimator for Hubble Compilation. The network architecture consists of an input layer with 31 z-values, four hidden layers with 22, 16, 11, and 8 neurons respectively (with ReLU activation), and an output layer with 6 cosmological parameters.}
	\label{fig:hubbleModel}
\end{figure*}

\subsubsection{H(z) Compilation:} The Hubble parameter dataset comprises 31 measurements spanning a redshift range of $z \in [0.07,1.965]$. These measurements were obtained using the Differential Age (DA) technique.  Table \ref{tab:hubbledata}  shows these measured \( H(z) \) values and associated uncertainties expressed in km s$^{-1}$ Mpc$^{-1}$ units~\citep{2002ApJ...573...37J}. 

\subsubsection{Pantheon Supernova Dataset:} A total of 1,048 Type Ia Supernovae measurements are contained within the Pantheon sample in the redshift range \( z \in [0.1,2.26] \)~\citep{1999ApJ...517..565P,1997AAS...191.8504P,2022ApJ...938..110B}

For both datasets, redshift points are arranged in ascending order to facilitate systematic generation of mock values for \( H(z) \) and \( d_L(z) \).

\begin{table}[htb]
\centering
\renewcommand{\arraystretch}{1.3} 
\setlength{\tabcolsep}{20pt} 
	\caption{Hubble parameter measurements $H(z)$ and respective uncertainties, $\sigma(H(z))$ at different redshifts $z$ ~\citep{2002ApJ...573...37J}.}
	\label{tab:hubbledata}
	\footnotesize 
	
	\begin{tabular}{lcc}
\toprule
		$z$ & $H(z)$ & $\sigma(H(z))$ \\\midrule
		0.07 & 69 & 19.6 \\ 
		0.09 & 69 & 12 \\ 
		0.12 & 68.6 & 26.2 \\ 
		0.17 & 83 & 8 \\ 
		0.1791 & 75 & 4 \\ 
		0.1993 & 75 & 5 \\ 
		0.2 & 72.9 & 29.6 \\ 
		0.27 & 77 & 14 \\  
		0.28 & 88.8 & 36.64 \\ 
		0.3519 & 83 & 14 \\ 
		0.3802 & 83 & 13.5 \\ 
		0.4 & 95 & 17 \\ 
		0.4004 & 77 & 10.2 \\ 
		0.4247 & 87.1 & 11.2 \\ 
		0.4497 & 92.8 & 12.9 \\ 
		0.47 & 89 & 34 \\ 
		0.4783 & 80.9 & 9 \\ 
		0.48 & 97& 62 \\ 
		0.5929 & 104 & 13 \\ 
		0.6797 & 92 & 8 \\ 
		0.7812 & 105 & 12 \\ 
		0.8754 & 125 & 17 \\ 
		0.88 & 90 & 40 \\ 
		0.9 & 117 & 23 \\ 
		1.0375 & 154 & 20 \\ 
		1.3 & 168 & 17 \\ 
		1.363 & 160 & 33.6 \\ 
		1.43 & 177 & 18 \\ 
		1.53 & 140 & 14 \\ 
		1.75 & 202 & 40 \\ 
		1.965 & 186.5 & 50.4 \\ 
		
		\bottomrule
	\end{tabular}
	\vspace{2mm}
	
\end{table}

\subsection{Data Simulations}

\subsubsection{H(z) Compilation:}

A uniform distribution is considered in the range \( \{50, 90\} \) km Mpc\textsuperscript{-1} sec\textsuperscript{-1} for \( H_0 \), the range \( \{0.1, 0.7\} \) for \( \Omega_{m0} \), and the range \( \{0.3, 0.9\} \) for \( \Omega_{\Lambda0} \). Using the numpy library's \texttt{random.uniform} function, we generate 120,000 random samples for each parameter, ensuring comprehensive coverage of the parameter space.

For each parameter set, we compute the Hubble parameter \( H(z) \) at 31 observed redshift points using Eqn.~\ref{eqn1}. This process yields 120,000 distinct realizations of \( H(z) \) measurements across the observed redshift range.

\subsubsection{Pantheon Supernova Dataset:} 

The Eqn.~\ref{eqn1} can be rewritten as

\begin{equation}
	H(z) = H_0 \sqrt{\Omega_m (1 + z)^3 + \Omega_\Lambda} ,
\end{equation}
here \( H_0 \) is fixed at 66.99 km s$^{-1}$ Mpc$^{-1}$

The luminosity distance \( d_L(z) \) is defined as:

\begin{equation}
	d_L(z) = (1 + z) \int_{0}^{z} \frac{c \, dz'}{H(z')} ,
\end{equation}
where \( c \) is the speed of light. The angular diameter distance \( d_A(z) \) can be calculated using the cosmic distance duality relation:

\begin{equation}
	d_A(z) = \frac{d_L(z)}{(1 + z)^2} .
\end{equation}
For Type Ia Supernovae (SNe Ia), the distance modulus \( \mu(z) \) is given by:

\begin{equation}
	\mu(z) = 5 \log_{10} \left( \frac{d_L(z)}{\text{Mpc}} \right) + 25 .
	\label{eq:dL_final}
\end{equation}

The numerical integration required for \( d_L(z) \) is performed using \texttt{Simpson's 1/3 method}, generating 120,000 realizations of \( \mu(z) \) at 1048 observed redshift points.

\subsection{Adding Noise to Data}
In supervised learning applications, training data should accurately reflect the statistical properties of test data—in this case, observational measurements. Each measurement $X$ is assumed to follow a Gaussian distribution N(X,\( \sigma \)\textsuperscript{2}), where $X$ represents the mean value and \( \sigma \)  denotes  associated measurement uncertainty. Yet, the values generated  by the theoretical model do not have errors. Thus, before training the artificial neural network (ANN), the simulated data should be adjusted to match the spread of the observational data.
We simulate observational uncertainties by adding Gaussian random noise to the synthetic datasets.

To achieve this, we incorporate   random noise following  Gaussian probability density to the simulated data, aligning it with the error level of the observational measurements used for cosmological parameter estimation. This adjustment helps to prevent distribution inconsistency, overfitting, and enhances the generalization of the trained model. For each measurement, random deviates are drawn from a normal distribution N(X,\( \sigma \)\textsuperscript{2}), where \( \sigma \) corresponds to the observed uncertainty at each redshift point, using the \texttt{random.normal} function. This random number is then added to both \( H(z) \) and \( \mu(z) \) corresponding to the data set. 

\subsection{Data Preprocessing}

The complete dataset is partitioned into three distinct subsets: a training set, a validation set, and a testing set. 10\textsuperscript{5} data points are allocated for training, 1.5 × 10\textsuperscript{4}  data points are allocated  for validation, and 5 × 10\textsuperscript{3}  data points for testing the CosmicANNEstimator predictions. This partitioning ensures robust model training, proper validation for hyperparameter optimization, and unbiased performance evaluation.

\begin{figure*}
	\centering
	\includegraphics[width=1\linewidth]{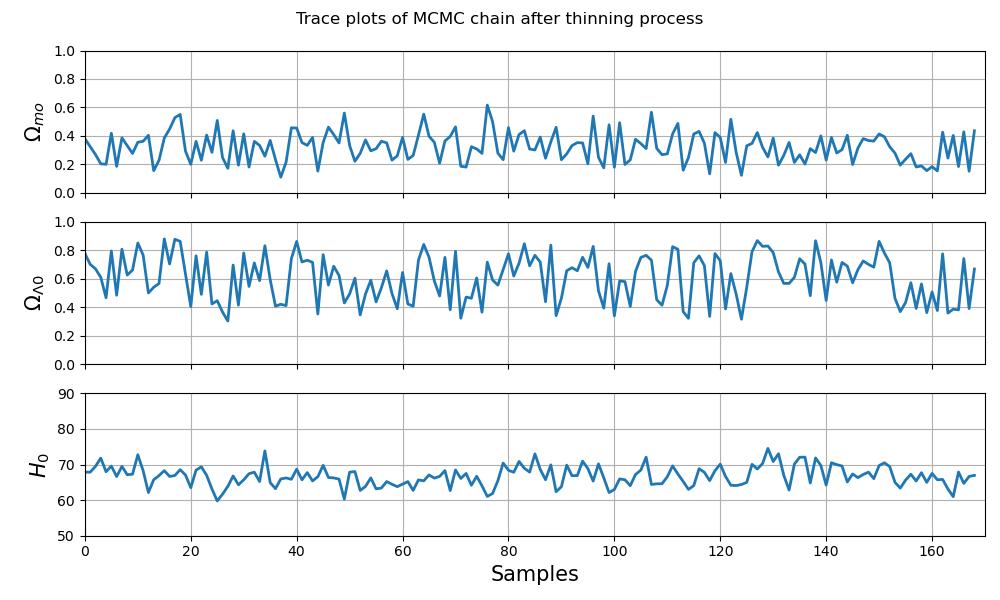}
	\caption{Figure showing trace plots of Hubble Parameter Data after performing thinning process}
	\label{HubbleTracePlots}
\end{figure*}

\begin{figure*}
	\centering
	\includegraphics[width=1\linewidth]{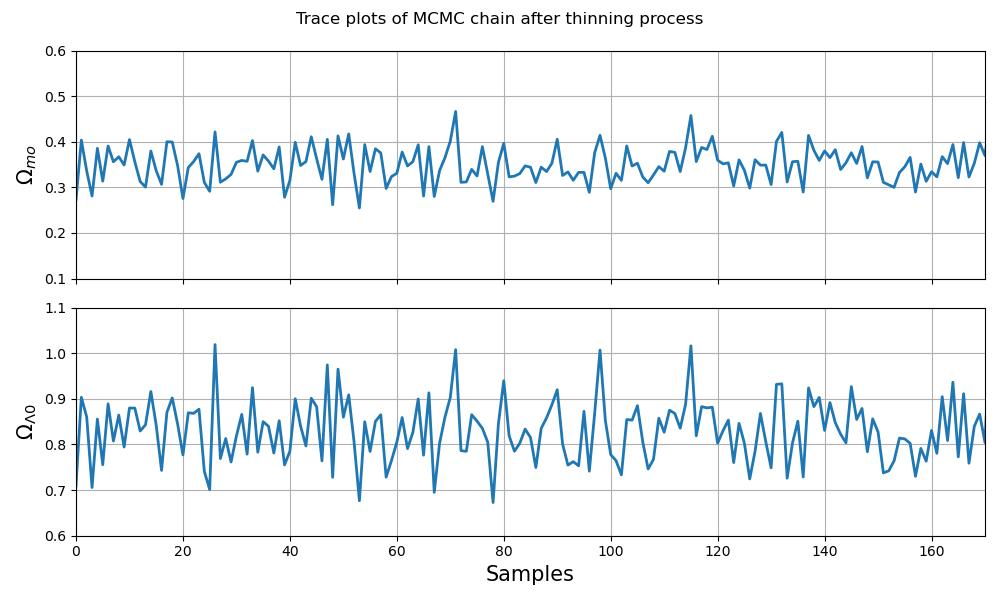}
	\  \caption{Figure showing trace plots of Supernova  Data after performing thinning process}
	\label{SupernovaTracePlot}
\end{figure*}

\subsubsection{Data Preprocessing for the Hubble Compilation:}For the Hubble parameter compilation, we implement a standardization procedure to optimize neural network training. We first compute two key statistical parameters from the noise-integrated measurements \(H(z)\) in our training dataset: the average value and its statistical spread (standard deviation). We then perform data standardization across all datasets (training, validation, and testing) using these computed parameters. We normalize each data by subtracting this average value and dividing by the standard deviation calculated from the training set. After this preprocessing, the standardized data is used for training the ANN.

\subsection{Proposed Model} \label{sec:propsed-model}
The core of our methodology centers on a sophisticated deep learning approach designed specifically for cosmological parameter estimation. We implement an artificial neural network (ANN) architecture that processes  measurements through a series of carefully designed transformations to predict fundamental cosmological parameters and their associated uncertainties.

Our parameter estimation framework employs a feed-forward neural network that systematically transforms observational data into cosmological parameter estimates. The network processes information through multiple stages, each implementing specific transformations designed to capture the complex relationships inherent in cosmological data. Observational measurements first enter the network through a dedicated input layer, where they undergo initial processing. These processed signals then propagate through multiple hidden layers, where non-linear transformations extract and refine relevant features from the data. Finally, a specialized output layer generates both parameter estimates and their corresponding uncertainties, providing a complete statistical description of the cosmological parameters of interest.

\subsubsection{Proposed Model for Hubble Compilation - CosmicANNEstimator-Hubble:}
For the analysis of Hubble parameter measurements, we implement a carefully optimized network architecture comprising \textbf{ six} distinct layers. The input layer contains 31 neurons, corresponding to the dimensionality of the Hubble parameter dataset. \textbf{This is followed by four  hidden layers implementing Rectified Linear Unit (ReLU) activation functions, with 22, 16, 11 and 8 neurons respectively}. The ReLU activation function,  expressed mathematically as f(x) = max(0,x), enables efficient network training while maintaining the capacity to capture non-linear relationships in the data. The architecture culminates in an output layer of 6 neurons, where the first three neurons generate direct parameter estimates for $H_0$, $\Omega_{m0}$, $\Omega_{\Lambda0}$,  while the remaining three neurons provide their corresponding uncertainty estimates. The network implements full connectivity between successive layers, ensuring comprehensive parameter space exploration through dense connections. 
For network optimization, we implement the Adaptive Moment Estimation (ADAM) algorithm, which combines the advantages of both adaptive gradient algorithms and momentum methods. The training process employs a carefully tuned learning rate of  5×10\textsuperscript{-4} , determined through extensive experimentation to ensure stable convergence while maintaining efficient parameter updates during backpropagation.~This optimization strategy allows the network to effectively navigate the parameter space while avoiding local minima and ensuring robust convergence to optimal weight and bias values.
The combination of noise-integrated training data, non-linear activation functions, and sophisticated optimization techniques enables our network to develop a robust understanding of the relationship between observational data and cosmological parameters. The training process continues until convergence criteria are met.~\textbf{Figure \ref{fig:hubbleModel} presents the architecture of CosmicANNEstimator for Hubble Compilation.}

\begin{figure*}[t!]
	\centering
	\includegraphics[width=1\linewidth]{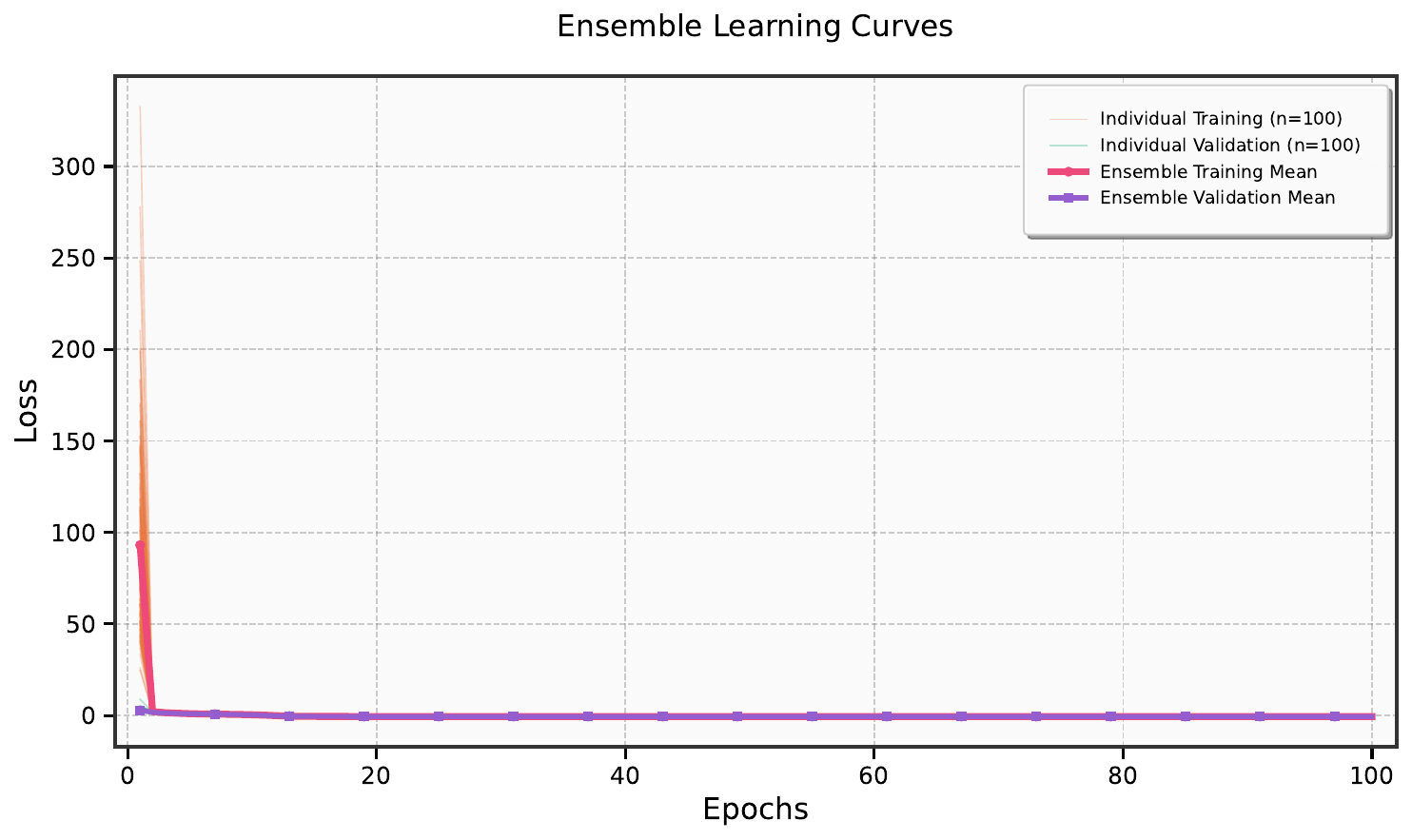}
	\caption{Figure showing Learning Curve of CosmicANNEstimator-Hubble Model}
	\label{fig2-hubblelossfn}
\end{figure*}

\begin{figure*}[t!]
	\centering
	\includegraphics[width=0.7\linewidth]{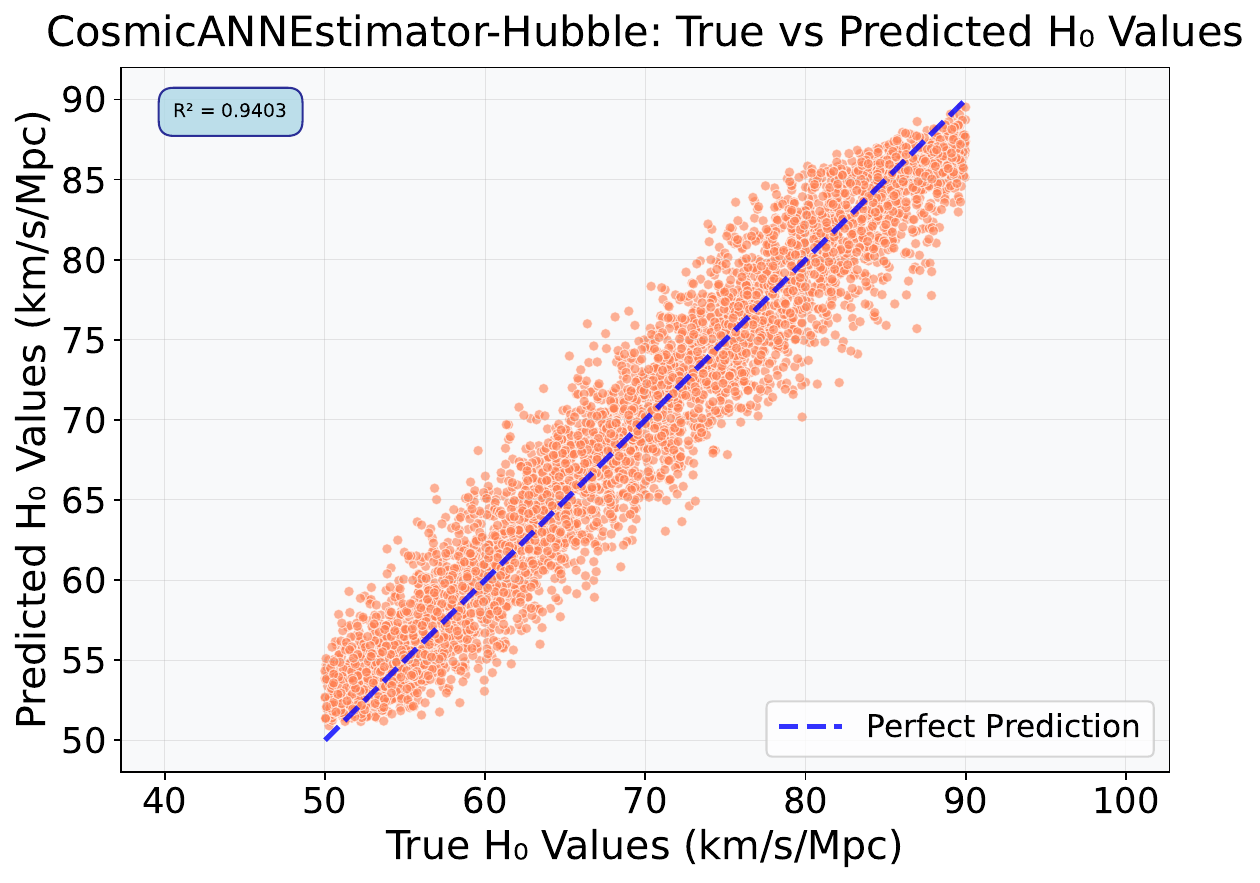}
	\caption{Figure showing Performance Evaluation of True vs 
		Predicted values for Hubble Constant ($H_0$) for CosmicANNEstimator-Hubble. Coefficient of determination, R²= 0.9403, shows  strong agreement between predicted and true $H_0$ values }
	\label{predictedvsActual}
\end{figure*}

\subsubsection{Proposed Model for Pantheon Supernova Dataset - CosmicANNEstimator-SN: }

~The implementation of CosmicANNEstimator for Type Ia supernova analysis employs a distinct architectural approach from the Hubble parameter estimation network. For supernova analysis, we focus exclusively on determining the density parameters ($\Omega_{m0}$, $\Omega_{\Lambda0}$) of the  $\Lambda$CDM model, working directly with uncalibrated supernova data without standardization preprocessing.

The supernova analysis network implements a deep architecture optimized for the high-dimensional nature of the Pantheon dataset. The input layer comprises 1048 neurons, corresponding to the dimensionality of the observational data. This is followed by a cascade of five hidden layers with progressively decreasing neuron counts: 512, 256, 128, 64, and 32 neurons, respectively with ReLU activation function. This pyramidal structure enables systematic feature extraction and dimensionality reduction while maintaining essential cosmological information. Full connectivity is maintained between successive layers to ensure comprehensive information flow through the network.

The network culminates in an output layer containing four neurons: two dedicated to parameter estimation ($\Omega_{m0}$ and $\Omega_{\Lambda0}$) and two for their associated uncertainties. This design differs from the Hubble parameter estimation network by excluding  $H_0$ prediction.

Network training implements specific hyperparameters optimized for the supernova analysis task. We employ a batch size of 128 samples, which balances computational efficiency with stable gradient estimates. The training proceeds for 200 epochs, allowing sufficient time for convergence while avoiding overfitting. A learning rate of 1×10\textsuperscript{-5} is implemented, enabling fine-grained weight adjustments necessary for the precise parameter estimation required in cosmological applications.

\begin{figure*}
	\centering
	\includegraphics[width=0.8\linewidth]{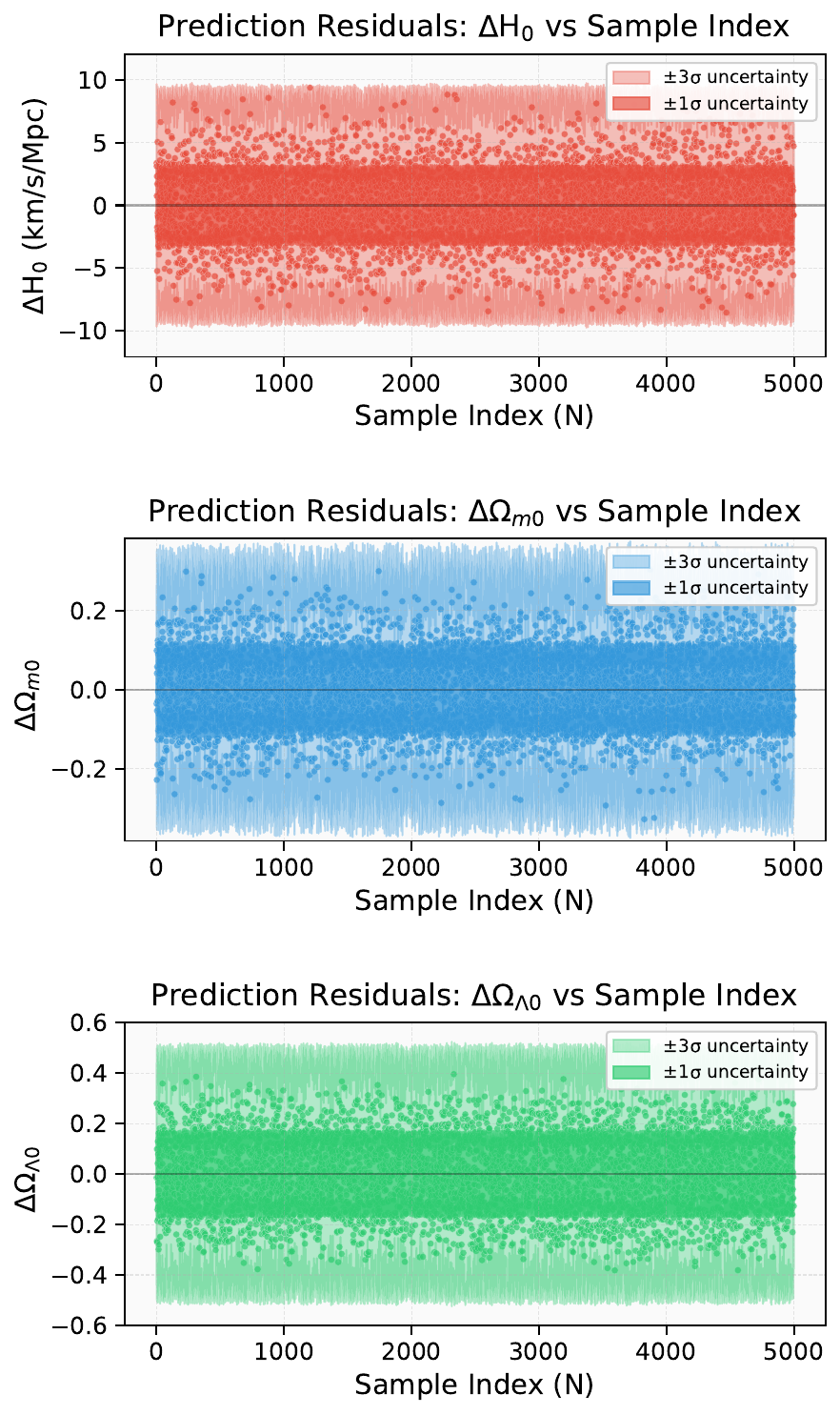}
	\caption{Figure showing distribution of differences between model outputs and true values of the Hubble constant, matter density and dark energy density parameters across various realizations (N) of the test set}
	\label{fig:predictnscomparison}
\end{figure*}

\subsection{Loss function}

Our methodology implements a heteroscedastic (HS)~\citep{2017arXiv170304977K} loss function for uncertainty quantification in parameter estimation. The function is defined as:
\begin{equation}
	L_{HS} = \frac{1}{2n} \sum_{q=1}^{n} \exp(-s_a)(y_a - \hat{y}_a)^2 + s_a
\end{equation}
where $n$ specifies the number of actual values. In the equation, $y_a$ and $\hat{y}_a$ represent the predictions and actual values respectively. Additionally, log variances ($\ln\sigma^2_a$)  corresponding to the predictions is represented by $s_a$  , where $\sigma_a$ specifies the aleatoric uncertainty of the prediction. By implementing this specific loss function design, we can determine the predictive uncertainty (aleatoric) associated with each individual output.

The output layer architecture of CosmicANNEstimator varies based on the dataset used. For Hubble Compilations, the output layer contains six neurons: three neurons estimate the cosmological parameters ($H_0$, $\Omega_{m0}$, $\Omega_{\Lambda0}$), while the other three calculate their corresponding uncertainties. In contrast, when processing the Pantheon Supernova Dataset, the output layer consists of four neurons: two neurons predict the cosmological density parameters ($\Omega_{m0}$, $\Omega_{\Lambda0}$), and the remaining two neurons compute their respective uncertainties.

The heteroscedastic loss function shares fundamental similarities with negative log-likelihood functions commonly employed in traditional MCMC-based cosmological parameter estimation. However, the optimization methodologies differ substantially. Traditional Bayesian approaches minimize noise variance through likelihood functions, typically assuming Gaussian distributions, to extract optimal parameter values from observational data.
In contrast, our neural network implementation represents a likelihood-evaluation-free inference method~\citep{2018MNRAS.477.2874A}, where the loss function quantifies the discrepancy between predicted and reference values based on training data. Despite these methodological differences, comparative analysis demonstrates that both approaches yield consistent results when applied to both Hubble measurements and Pantheon Supernova data, validating the robustness of our neural network methodology.

\subsection{Model Training} \label{sec:model-training}

The CosmicANNEstimator adopts an ensemble learning strategy encompassing 100 independent training realizations with standardized hyperparameter configurations. Each training realization employs a validation set containing  1.5 × 10\textsuperscript{4} data points for performance optimization. 
\textbf{These 100 trained models are subsequently employed for cosmological parameter estimation from observational datasets. }

The training process demonstrates robust convergence characteristics, with no evidence of overfitting or underfitting across the ensemble. The complete training procedure for the 100-member ensemble for CosmicANNEstimator-Hubble model required approximately 180 minutes of computation time on an AMD64 Family 25 Model 80 CPU system (8 cores, 16 threads, 3.2 GHz processor).

\subsection{Model prediction on Simulated Data}
\textbf{The parameter estimation procedure utilizes a test set containing 5 × 10\textsuperscript{4}  data points, evaluated across the 100-member ensemble of trained models. Each of the 100 ensemble model generates both parameter estimates and log-variances through the heteroscedastic loss function.}

\textbf{Final parameter estimates are computed as the mean of predictions across all ensemble members for each test realization. Uncertainty calculations follow a systematic two-step process: first, log-variances from each ensemble member are exponentiated to obtain individual variances, then averaged across the 100 members. Final uncertainties are derived by taking the square root of these averaged variances.}

\textbf{The residuals between predicted and actual values are evaluated to verify that parameter differentials predominantly remain within three times the predicted uncertainty ($\pm 3\sigma$) bounds, thereby confirming robust agreement between predictions and reference values and validating the model's ability to provide estimates with accurate uncertainty quantification across the parameter space.}

\subsection{MCMC}\label{sec:parallel-mcmc}
We implement the Metropolis-Hastings algorithm~\citep{hastings1970} to sample posterior distributions of three cosmological parameters ($H_0$, $\Omega_{m0}$, and $\Omega_{\Lambda0}$), utilizing identical prior ranges as the CosmicANNEstimator method. The likelihood function for the  data  is defined as:

\begin{equation}
	\mathcal{L}_{data}(\theta) = e^{- \chi_{data}^2(\theta)\big /2}
	\label{liklie}
\end{equation}

where $\chi_{\rm data}^2$ denotes Hubble data or Supernova data based on the dataset and  $\theta$ represents the model's variable parameters. The optimal parameter values ($\theta$) are determined by identifying those that yield the lowest possible $\chi^2$ value.

We employed MCMC sampling to identify high-confidence regions for our model parameters, constrained by the observational dataset. Through likelihood analysis, we minimized the $\chi^2$ function (Eq. \ref{liklie}), leading to the determination of optimal model parameters at the $\chi^2$ minimum. 

We initiated our MCMC analysis using a conventional proposal function:
$ \theta^j_{i+1}  = \theta^j_i  + \delta\theta^j \eta^j$
where $j$ denotes the parameter index, $\delta^j$ represents the predefined rms step size, and $\eta^j$ is a Gaussian stochastic variate with zero mean and unit variance. Given that our model incorporates three parameters with strong intercorrelations, this basic approach proved suboptimal. Therefore, we implemented several optimization techniques on the MCMC chain samples to enhance both the acceptance ratio and chain convergence.
Following the generation of a preliminary chain with sufficient samples, we evaluated the chain's autocorrelation. This analysis provided crucial insights into determining the necessary number of Markov chain iterations required for effectively independent samples. Subsequently, we applied thinning to the initial chain to reduce sample correlations, thereby decreasing the overall sample size while maintaining statistical validity.

From the resulting samples, we calculate the covariance matrix:
$C_{ij} = \big<\delta\theta^i\delta\theta^j\big>$

We then apply Cholesky decomposition to this matrix:
$\boldsymbol C =\boldsymbol L \boldsymbol L^{t}$
where $\boldsymbol C$ represents the covariance matrix, $\boldsymbol L$ denotes the lower triangular matrix, and $\boldsymbol L^t$ is its conjugate transpose. Using these results, we reformulate our proposal function as:
$ \theta_{i+1}  = \theta_i  + \alpha \boldsymbol L \boldsymbol\eta$
Here, $\boldsymbol\eta$ represents a vector of Gaussian variates, and $\alpha$ is an overall scale factor, initially set to approximately $0.3$. This reformulation ensures that proposed samples exhibit the appropriate covariance structure, substantially improving sampling efficiency.
To maintain optimal sampling performance, we implement bounds on the acceptance ratio, requiring it to fall between $5\%$ and $80\%$. The scale factor $\alpha$ can be adjusted when these bounds are exceeded, preventing either excessively large or small step sizes.

Trace plot analysis presented in Figures \ref{HubbleTracePlots} and \ref{SupernovaTracePlot}, offers additional confirmation of proper chain convergence through post-thinning visualization of parameter evolution. The Hubble parameter trace plots exhibit stable random fluctuations around mean values for all three parameters ($H_0$, $\Omega_{m0}$, $\Omega_{\Lambda0}$), while the supernova analysis demonstrates similar stability for its two parameters ( $\Omega_{m0}$, $\Omega_{\Lambda0}$). In both cases, the trace plots show no evidence of systematic trends or long-term drifts, displaying instead consistent oscillation patterns that indicate proper exploration of the parameter space. The absence of chain sticking or extreme parameter excursions, combined with uniform exploration around central values, provides strong evidence for reliable posterior sampling in both analyses.

\subsubsection{Convergence test: }
The goal of MCMC sampling is to draw samples from a target distribution, ideally reaching a point where samples are representative of this distribution. However, MCMC chains can take time to converge, especially if initialized in different regions of the parameter space. The Gelman-Rubin diagnostic provides a quantitative way to evaluate whether the chains have "mixed" well enough and converged to the target distribution, ensuring that further sampling will yield reliable estimates. The statistical relationships can be written mathematically as:
\begin{equation}
	E = \frac{1}{M(N-1)} \sum_{i=1}^{M} \sum_{j=1}^{N} (\Theta_{ij} - \bar{\Theta}_i)^2
\end{equation}
\begin{equation}
	F = \frac{N}{M-1} \sum_{i=1}^{M} (\bar{\Theta}_i - \bar{\Theta})^2
\end{equation}
\begin{equation}
	G = (1-\frac{1}{N})E + \frac{1}{N}F
\end{equation}
In these equations, the variable $M$ indicates the total chain count, while $N$ signifies the number of data points within each individual chain. We evaluate the MCMC algorithm's convergence using the R statistic developed by Gelman and Rubin, R = V/W, which should approach unity for proper convergence. Our analysis yields R-values of 0.9996 and 0.995 for the Hubble parameter and supernova data analyses, respectively, indicating excellent convergence.

\section{Results and Analysis}
We evaluate two distinct implementations of CosmicANNEstimator: one optimized for Hubble measurements(CosmicANNEstimator-Hubble) and another designed for Type Ia Supernova (SnIa) observations(CosmicANNEstimator-SN). Each variant undergoes testing with its corresponding dataset—the Hubble-based model with Hubble parameter measurements and the Supernova-based model with Pantheon Supernova observations. The models' predictions are validated against estimations from dedicated MCMC analyses for their respective datasets.

\subsection{ Predictions for Hubble Compilation using CosmicANNEstimator-Hubble}
\subsubsection{Predictions on Test Set:}

The learning dynamics of the Hubble parameter model demonstrate robust convergence characteristics, as illustrated in Figure~\ref{fig2-hubblelossfn}. 
\textbf{The ensemble training (red curve) and validation (purple curve) losses demonstrate  convergence within the first 10-12 epochs, maintaining minimal separation throughout the 100-epoch training period. The ensemble means for both training and validation losses remain consistently low and closely aligned, indicating optimal model generalization without overfitting across all 100 ensemble members.}

We evaluate the model's  accuracy by testing it against H(z) measurements from our reserved test dataset to estimate the Hubble constant ($H_0$), matter density ($\Omega_{m0}$), and dark energy density ($\Omega_{\Lambda0}$) . Figure  \ref{predictedvsActual} demonstrates the correlation between predicted and actual Hubble constant values, revealing a strong linear relationship that confirms the model's estimation accuracy.

Parameter prediction accuracy is quantified through the deviation metric:
\[
\Delta y = \hat{y} - y
\]
where $y$ denotes the model predictions and $\hat{y}$ represents the reference (true) values.

\textbf{Figure \ref{fig:predictnscomparison} presents a comprehensive analysis of parameter prediction residuals across the test set of 5000 samples. The three panels display the deviations for the Hubble constant $\Delta H_0$ (top), matter density parameter $\Delta\Omega_{m0}$ (middle), and dark energy density parameter $\Delta\Omega_{\Lambda0}$ (bottom). Each panel includes shaded bands representing one and three times the predicted uncertainty ($\pm 1\sigma$ and $\pm 3\sigma$) for the respective parameters. The analysis reveals that parameter residuals are predominantly contained within the $\pm 3\sigma$ uncertainty bounds, with the vast majority falling within the $\pm 1\sigma$ bands, demonstrating excellent agreement between model predictions and reference values. The distribution of residuals across all parameters exhibits symmetric behavior around zero. This confirms the model's ability to deliver accurate predictions with reliable uncertainty quantification.}



\subsubsection{Predictions On Observational Data: }

The CosmicANNEstimator, trained on simulated $H(z)$ values incorporating observational noise patterns, processes standardized observed $H(z)$ measurements to extract cosmological parameters and their uncertainties. A parallel MCMC analysis as mentioned in \hyperref[sec:parallel-mcmc]{Section 2.9} provides independent parameter estimations using identical observational data, enabling direct comparison of both methodologies.

Our analysis yields consistent parameter estimates across both methods for three fundamental cosmological parameters:

\begin{table}[htb]
	\centering
	\caption{Comparison of parameter values $H_0$ (in km\,s$^{-1}$\,Mpc$^{-1}$), $\Omega_{m0}$, and $\Omega_{\Lambda0}$ estimated by MCMC and CosmicANNEstimator for Hubble data~\citep{2002ApJ...573...37J}.}
	\label{Hubblecomparison}
	\begin{tabular}{lcc}
		\toprule
		\textbf{Parameter} & \textbf{MCMC} & \textbf{CosmicANNEstimator} \\
		\midrule
		$H_0$ & $66.99 \pm 2.99$ & $67.14 \pm 3.12$ \\
		$\Omega_{m0}$ & $0.321 \pm 0.104$ & $0.317 \pm 0.109$ \\
		$\Omega_{\Lambda0}$ & $0.620 \pm 0.154$ & $0.621 \pm 0.165$ \\
		\bottomrule
	\end{tabular}
	
	\vspace{2mm}
	\footnotesize{The uncertainties are expressed as $\pm 1\sigma$.}
\end{table}

The comparative analysis reveals strong concordance between both methods across all three fundamental cosmological parameters \textbf{ as mentioned in Table~\ref{Hubblecomparison}}. For the Hubble constant ($H_0$), CosmicANNEstimator yields $H_0 = 67.1448 \pm 3.122 \, \text{km s}^{-1} \, \text{Mpc}^{-1}$, closely matching the MCMC estimate of $H_0 = 66.9943 \pm 2.9918 \, \text{km s}^{-1} \, \text{Mpc}^{-1}$. The matter density parameter ($\Omega_{m0}$) shows similar agreement, with CosmicANNEstimator producing $\Omega_{m0} =0.3174 \pm 0.1087$ compared to the MCMC value of $\Omega_{m0} = 0.3213 \pm 0.1038$. For the dark energy density parameter ($\Omega_{\Lambda0}$), CosmicANNEstimator estimates $\Omega_{\Lambda0} =0.6209\pm 0.1654$, in strong agreement with the MCMC result of $\Omega_{\Lambda0} = 0.6204 \pm 0.1537$.

\begin{figure*}[t!]
\centering
\includegraphics[width=.9\linewidth]{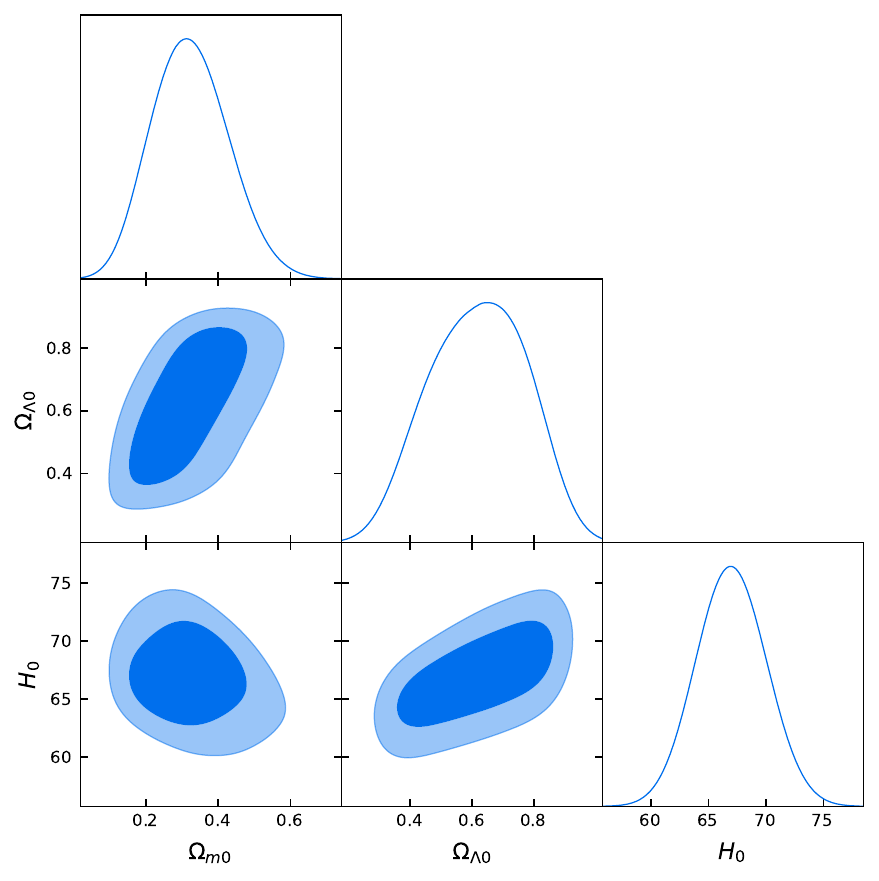}
\caption{ Figure showing MCMC posterior distributions for cosmological parameters for Hubble dataset}
\label{HubbleMCMC}
\end{figure*}

The MCMC posterior distributions, illustrated in Figure \ref{HubbleMCMC} through a corner plot, provide deeper insight into parameter relationships. The marginalized distributions show peaks consistent with our point estimates: 67-68  km Mpc$^{-1}$ sec$^{-1}$,$\Omega_{\Lambda0}$ centered near 0.6, and $\Omega_{m0}$ approximately at 0.3.  The two-dimensional joint distributions, depicted with 68\% (dark blue) and 95\% (light blue) confidence contours, reveal the correlations between parameters and confirm the robustness of our estimates.

The consistency between CosmicANNEstimator and MCMC results, both in parameter values and their associated uncertainties, validates the effectiveness of our neural network approach for cosmological parameter estimation of Hubble Observations.

\subsection{Predictions for  Pantheon Supernova Dataset using CosmicANNEstimator-SN}
\subsubsection{Predictions on Test Set:}
The CosmicANNEstimator's predictive capabilities for Type Ia Supernovae data are evaluated using noise-incorporated \( d_L(z) \) measurements from the test dataset. The model demonstrates robust performance in estimating both matter density (\( \Omega_{m0} \)) and dark energy density (\( \Omega_{\Lambda0} \)) parameters, with predictions exhibiting strong alignment with target values.

\begin{figure*}[t!]
\centering
\includegraphics[width=1\linewidth]{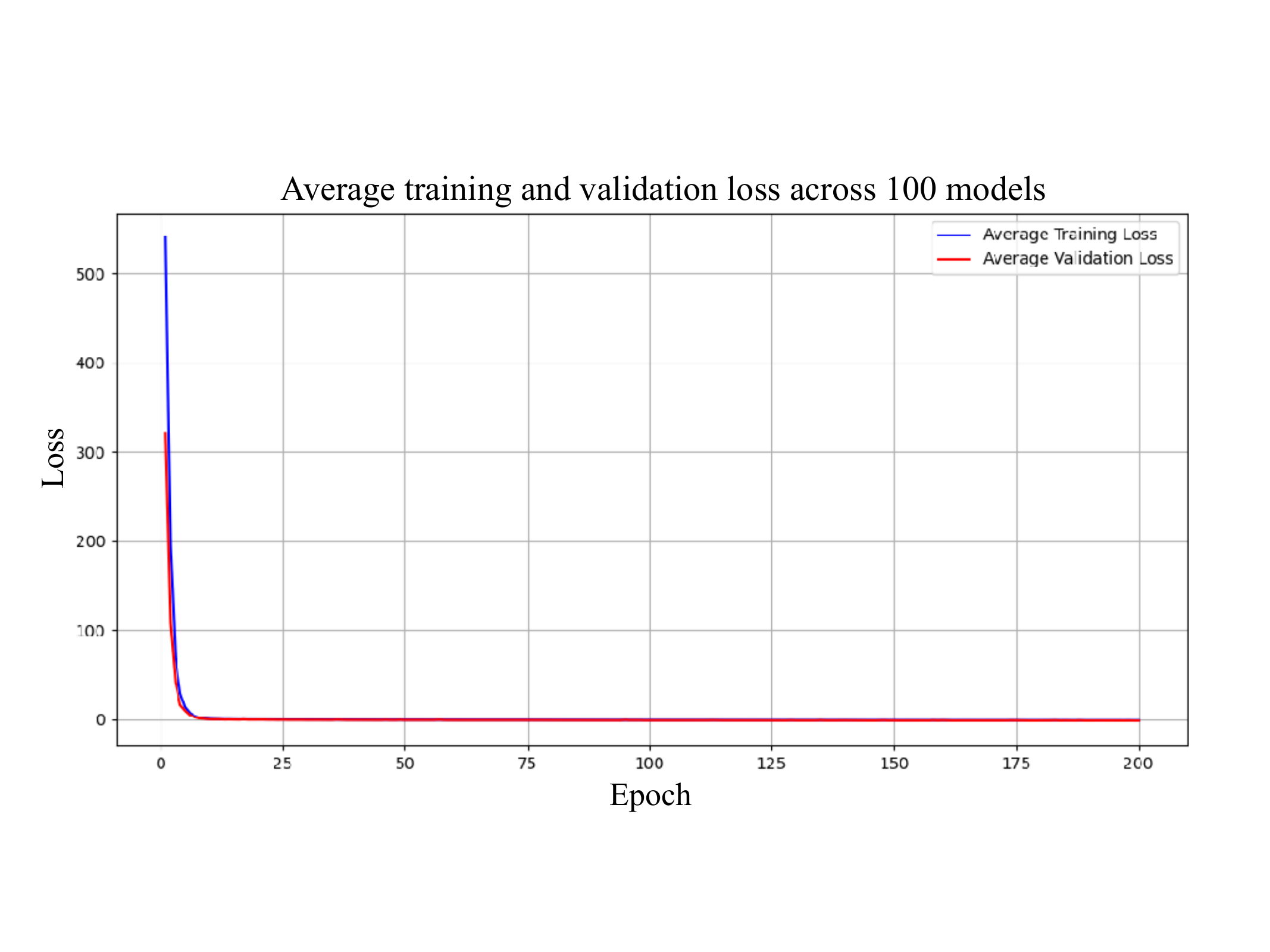}
\vspace{-2.0cm} 
\caption{Figure showing learning curve of CosmicANNEstimator-SN Model}
\label{SNLossFn}
\end{figure*}

Analysis of the training dynamics reveals  learning behavior across epochs, as shown in Figure \ref{SNLossFn}.  At the beginning (around epoch 0 to 20), there is a significant drop in both training and validation loss, indicating that the model quickly learns patterns in the data in the early epochs. After around epoch 20, both the training and validation loss values stabilize at a much lower value, and the decrease in loss becomes minimal for the remainder of the training.The average training loss (in blue) and validation loss (in red) closely follow each other throughout the epochs, with minimal divergence.This suggests that the model has a good fit and is not overfitting, as there isn’t a large gap between training and validation losses.

\subsubsection{Predictions On Observational Data: }

The CosmicANNEstimator for Type Ia supernova analysis employs simulated luminosity distance measurements that incorporate uncertainties matching the observed Pantheon dataset characteristics. We process observed luminosity distance values through the trained model to estimate two fundamental cosmological parameters: matter density ($\Omega_{m0}$) and dark energy density ($\Omega_{\Lambda0}$), along with their associated uncertainties. A parallel MCMC analysis following methodology detailed in \hyperref[sec:parallel-mcmc]{Section 2.9} provides independent parameter estimations using identical Pantheon supernova measurements. Table \ref{SNCOMPARISON} presents the comprehensive comparison between both approaches.

For the matter density parameter( $\Omega_{m0}$), MCMC yields a value of $0.35006 \pm 0.04002$, while the CosmicANNEstimator estimates $ 0.2830 \pm 0.0588$. For the dark energy density parameter $\Omega_{\Lambda0}$, MCMC calculates $0.8308 \pm 0.0677$, compared to CosmicANNEstimator's estimation of $0.9025 \pm 0.1089$. Although the best-fit values of parameters ($\Omega_{m0}$ and $\Omega_{\Lambda0}$)  from supernova datasets deviate slightly from their MCMC values, they are consistent at 1-sigma with their corresponding MCMC results. It is also interesting to  note that the ANN method produces larger uncertainties compared to MCMC.

\begin{table}[htb]
	\centering
	\caption{Comparison of parameter values $\Omega_{m0}$ and $\Omega_{\Lambda0}$ estimated by MCMC and CosmicANNEstimator for Supernova data.}
	\label{SNCOMPARISON}
	\begin{tabular}{lcc}
		\toprule
		\textbf{Parameter} & \textbf{MCMC} & \textbf{CosmicANNEstimator} \\
		\midrule
		$\Omega_{m0}$ & $0.35006 \pm 0.04002$ & $0.2830 \pm 0.0588$ \\
		$\Omega_{\Lambda0}$ & $0.8308 \pm 0.0677$ & $0.9025 \pm 0.1089$ \\
		\bottomrule
	\end{tabular}
	
	\vspace{2mm}
	\footnotesize{The uncertainties are expressed as $\pm 1\sigma$.}
\end{table}

\begin{figure*}[th!]
\centering
\includegraphics[width=1\linewidth]{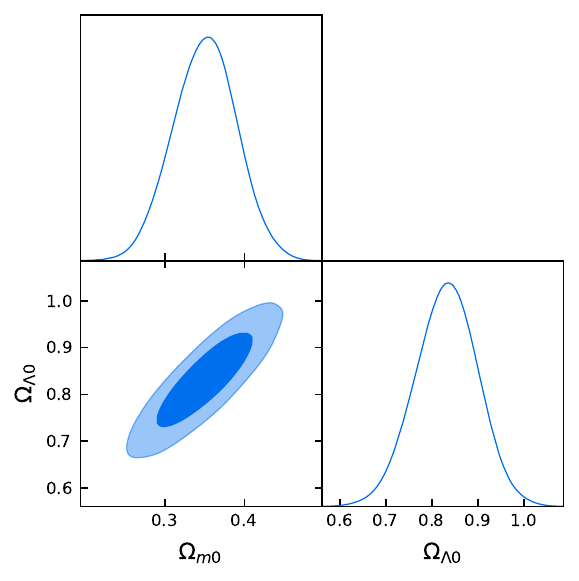}
\caption{Figure showing MCMC posterior distributions for cosmological parameters for Supernova Dataset}
\label{SNMCMCl}
\end{figure*}

The figure \ref{SNMCMCl} illustrates the MCMC posterior distributions for cosmological parameters derived from Type Ia Supernova data analysis. The plot is structured as a corner plot showing both marginalized and joint distributions for matter density ($\Omega_{m0}$) and dark energy density ($\Omega_{\Lambda0}$) parameters. The diagonal panels display one-dimensional marginalized posterior distributions, with $\Omega_{m0}$ centered approximately at 0.35 and $\Omega_{\Lambda0}$ peaking around 0.83. The off-diagonal panel reveals the two-dimensional joint distribution with confidence intervals represented by dark blue (68\%) and light blue (95\%) contours. The narrow width of the posterior distributions suggests well-constrained parameter estimates from the Supernova data.

Using Hubble parameter data we obtain the present value of Hubble Constant, $H_0 =67.1448 \pm 3.122$ km s$^{-1}$ Mpc$^{-1}$, demonstrating remarkable agreement with the latest Planck 2018 measurements ~\citep{2020A&A...641A...6P}.

\section{Effect of Hyperparameters}

Neural network performance is inherently sensitive to architectural and training configuration choices. While default hyperparameter values were established in the preceding analysis \hyperref[sec:propsed-model]{Section 2.5}, systematic investigation of their impact on parameter estimation accuracy is essential for optimal model design. This section presents a comprehensive sensitivity analysis examining how  hyperparameters---including network depth, activation function selection and learning rate influence on the results for Hubble CosmicANNEstimator model.

To evaluate the influence of model design and training choices on predictive performance, we define a significance metric to quantify deviations from benchmark results. For each cosmological parameter $\theta \in \{H_0, \Omega_{m0}, \Omega_{\Lambda0}\}$ the neural network provides both a mean prediction and an associated uncertainty ($\sigma$). 

The absolute deviation between ANN predictions and MCMC reference values is calculated as:
\begin{figure*}
	\centering
	\includegraphics[width=0.7\linewidth]{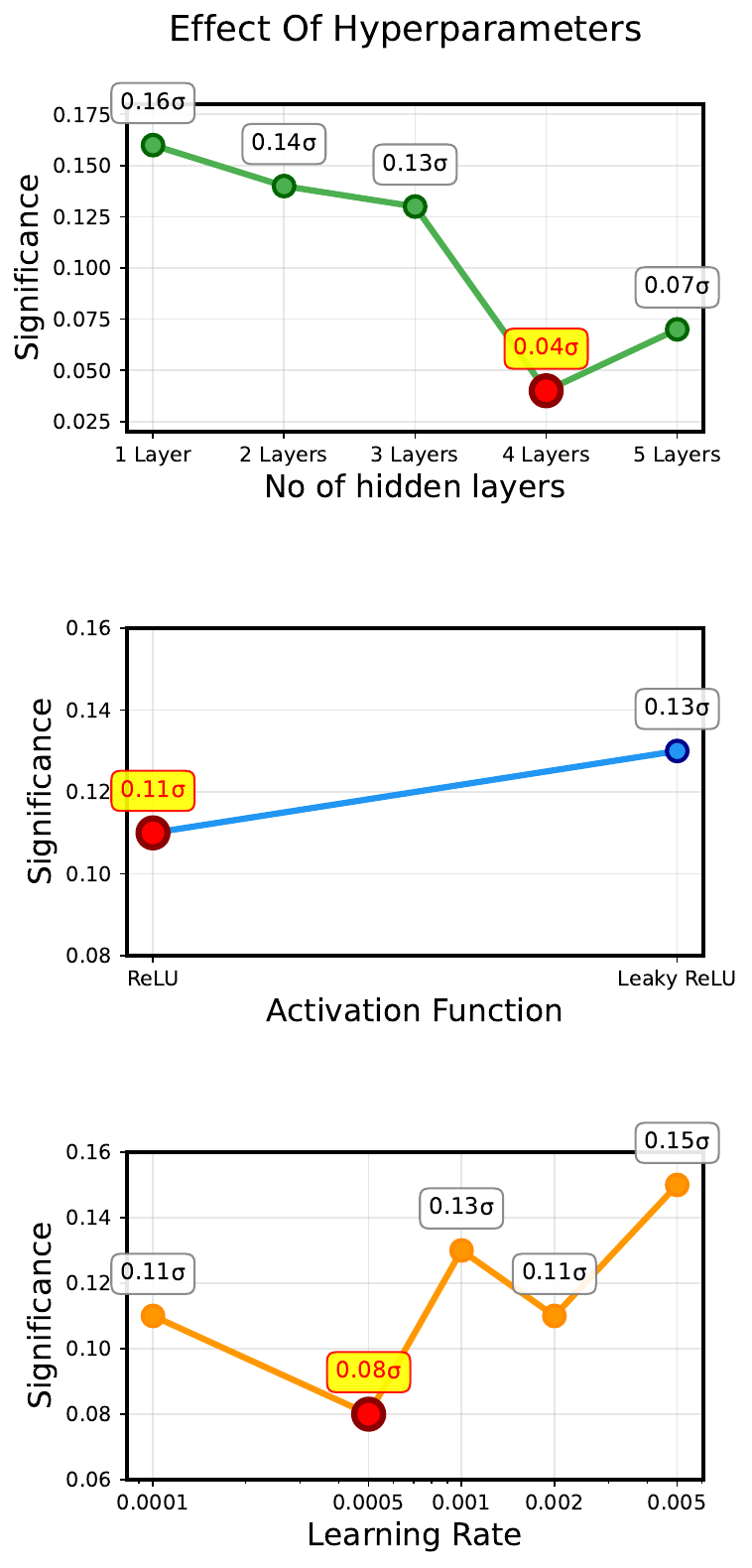}
	\caption{Figure showing hyperparameter sensitivity analysis for CosmicANNEstimator-Hubble with significance levels of ANN-MCMC deviations }
	
	\label{fig:hyperparameter-effects}
\end{figure*}

\begin{equation}
	\Delta\theta = |\theta^{\text{ANN}} - \theta^{\text{MCMC}}|
\end{equation}

This deviation is then normalized by the model's predicted uncertainty:

\begin{equation}
	\delta\theta = \frac{\Delta\theta}{\sigma_\theta}
	\label{deviation}
\end{equation}

which expresses how many standard deviations the ANN prediction lies away from the MCMC benchmark. The overall significance is defined as the sum of the relative deviations across all parameters:

\begin{equation}
	S = \delta H_0 + \delta\Omega_{m0} + \delta\Omega_{\Lambda0}
\end{equation}

This metric serves as a consolidated measure of how closely the ANN predictions align with MCMC results, with lower significance values indicating stronger overall consistency across all cosmological parameters.

\subsection{The Number of Hidden Layers}

We first test the effect of the number of hidden layers in the ANN model. We design five different ANN structures with the number of hidden layers from 1 to 5, where the number of neurons in each layer is set according to Equation (\ref{eq:neurons_per_layer}). 

\begin{equation}
	N_i = \left\lfloor \frac{N_{\text{in}}}{F^i} \right\rfloor, \quad i = 1, 2, \ldots, n_{\text{layers}}
	\label{eq:neurons_per_layer}
\end{equation} 

The decreasing factor $F$ is calculated as:

\begin{equation}
	F = \left(\frac{N_{\text{in}}}{N_{\text{out}}}\right)^{\frac{1}{n_{\text{layers}} + 1}}
	\label{eq:decreasing_factor}
\end{equation}

where:
\begin{itemize}
	\item $N_{\text{in}}$ is the number of input neurons
	\item $N_{\text{out}}$ is the number of output neurons  
	\item $n_{\text{layers}}$ is the number of hidden layers
\end{itemize}

where $\lfloor \cdot \rfloor$ denotes the floor function to ensure integer neuron counts.

In addition, the activation function is ReLU, and the learning rate is 0.0005. Then, five sets of ANNs are trained, and the parameters are predicted using these models according to the procedure in \hyperref[sec:model-training]{Section 2.7}. Furthermore, we calculate significance of deviations between ANN and MCMC results (Equation \ref{deviation}). The significance as a function of the number of hidden layers is shown in the top panel of Figure \ref{fig:hyperparameter-effects}, where the maximum deviation is $0.16\sigma$ (for the ANN with one hidden layer) and the minimum deviation is $0.04\sigma$ (for the ANN with four hidden layers). 

Starting with a single hidden layer, the network exhibits moderate performance with a significance of 0.16$\sigma$. This relatively high deviation suggests that a shallow architecture lacks sufficient representational capacity to capture the complex non-linear relationships inherent in cosmological parameter estimation. As the network depth increases to 2 layers, performance improves marginally to 0.14$\sigma$, followed by a further enhancement at 3 layers (0.13$\sigma$). This gradual improvement indicates that additional layers initially provide beneficial feature abstraction capabilities, allowing the network to learn hierarchical representations of the input cosmological data. The most significant performance gain occurs at 4 hidden layers, where the significance drops  to 0.04$\sigma$. This represents the optimal architecture depth, achieving excellent agreement with MCMC benchmark results. The substantial improvement suggests that 4 layers provide the ideal balance between model complexity and generalization capability, enabling effective extraction of cosmological parameter relationships without overfitting. However, further increasing the depth to 5 layers results in performance degradation, with significance rising to 0.07$\sigma$. Therefore, we adopt the ANN structure that has four hidden layers in our analysis.

\subsection{Activation Function}

To test the effect of activation function on the results of parameter estimations, we select two kinds of rectified units: rectified linear (ReLU) and leaky rectified linear (LeakyReLU).

The ReLU activation function is defined as:
\begin{equation}
	f(x) = \begin{cases}
		x & \text{if } x > 0 \\
		0 & \text{if } x \leq 0
	\end{cases}
\end{equation}
ReLU can introduce non-linearity while maintaining gradient flow for positive inputs. 
The leaky ReLU is defined as:
\begin{equation}
	f(x) = \begin{cases}
		x & \text{if } x > 0 \\
		\alpha x & \text{if } x \leq 0
	\end{cases}
\end{equation}
where $\alpha$ is a small positive constant, we set $\alpha$ to be 0.01 in our analysis.

Leaky ReLU is designed to address the ``dying ReLU'' problem by allowing small negative gradients.

In our analysis, the structure of the ANN with four hidden layers is adopted,  and
the learning rate is 0.0005. With the same procedure as \hyperref[sec:model-training]{Section 2.7}, we estimate cosmological parameters with ANNs by adopting these two different activation functions. After obtaining parameters, the mean deviations of parameters between the ANN results and MCMC values are calculated along with the significance, shown in the central panel of Figure \ref{fig:hyperparameter-effects}. The standard ReLU function demonstrates superior performance with a significance of 0.11$\sigma$, outperforming Leaky ReLU which achieves 0.13$\sigma$. Therefore, the ReLU activation function is used in this work.

\subsection{Learning Rate}

The learning rate analysis, depicted in the bottom panel of Figure \ref{fig:hyperparameter-effects}, reveals strong sensitivity to this hyperparameter across five different values ranging from 1×10$^{-4}$ to 5×10$^{-3}$.

At the lowest learning rate of 1×10$^{-4}$, the model achieves moderate performance with a significance of 0.11$\sigma$. The modest performance suggests that while training stability is maintained, the optimization process may not effectively explore the parameter space to find optimal network weights. The optimal performance occurs at a learning rate of 5×10$^{-4}$, yielding the lowest significance of 0.08$\sigma$. This intermediate value appears to strike the ideal balance between training stability and convergence efficiency. The learning rate is sufficiently large to enable effective gradient descent while remaining small enough to prevent overshooting optimal solutions. As the learning rate increases to 1×10$^{-3}$, performance degrades significantly to 0.13$\sigma$. Further increases to 2×10$^{-3}$ and 5×10$^{-3}$ result in continued deterioration, with significances of 0.11$\sigma$ and 0.15$\sigma$ respectively. 

\subsection{Effect of hyper parameters on CosmicANNEstimator-SN}
A similar hyperparameter optimization study was conducted for the supernova model architecture, following the same systematic approach employed for the Hubble Model analysis. The investigation revealed that a 5-layer neural network configuration with ReLU activation functions provided optimal performance for supernova parameter estimation tasks. Notably, the supernova model required a more conservative learning rate of 1×10$^{-5}$, an order of magnitude lower than the Hubble model's optimal rate of 5×10$^{-4}$. This reduced learning rate was essential to prevent instabilities during training while maintaining convergence to parameter estimates. 

\section{Conclusion}

We have developed a comprehensive methodology for cosmological parameter estimation using artificial neural networks through the CosmicANNEstimator framework. Our approach implements two specialized architectures: CosmicANNEstimator-Hubble and CosmicANNEstimator-SN, each specifically designed and optimized for analyzing Hubble parameter measurements and Type Ia Supernova datasets, respectively. The framework's training implementation utilized $10^5$ mock samples of $H(z)$ and $d_L(z)$ values, incorporating realistic noise characteristics that match observational data properties. The models employ distinct prior ranges tailored to their respective datasets: CosmicANNEstimator-Hubble uses $H_0 \in \{50, 90\}$ km Mpc$^{-1}$ s$^{-1}$, $\Omega_{m0} \in \{0.1, 0.7\}$, and $\Omega_{\Lambda 0} \in \{0.3, 0.9\}$, while CosmicANNEstimator-SN employs $\Omega_{m0} \in \{0, 1\}$ and $\Omega_{\Lambda 0} \in \{0, 1.5\}$.

CosmicANNEstimator-Hubble features a \textbf{ six-layer} architecture with an input layer of 31 neurons, \textbf{ four} hidden layers with ReLU activation functions  (\textbf{ 22, 16, 11, and 8 }neurons respectively), and an output layer predicting six parameters ($H_0$, $\Omega_{m0}$, $\Omega_{\Lambda 0}$) with their associated uncertainties. CosmicANNEstimator-SN employs a deeper seven-layer architecture, beginning with 1048 input neurons corresponding to observational data points, followed by five hidden layers with progressively decreasing neuron counts (512, 256, 128, 64, and 32 neurons), concluding with an output layer predicting two parameters ($\Omega_{m0}$, $\Omega_{\Lambda 0}$) and their uncertainties. Both models underwent rigorous validation using $1.5 \times 10^4$ mock data samples and comprehensive testing with $5 \times 10^3$ samples. Testing results demonstrate that differences between target values and predictions consistently fall within three standard deviations of the predicted uncertainties.

The trained models were successfully applied to real observational datasets. CosmicANNEstimator-Hubble analyzed 31 observed Hubble parameter measurements spanning redshift range $0.07 \leq z \leq 1.965$, yielding $H_0 = \textbf{ 67.1448} \pm \textbf{ 3.122}$ km Mpc$^{-1}$ s$^{-1}$, $\Omega_{m0} =\textbf{  0.3174} \pm \textbf{ 0.1087}$, and $\Omega_{\Lambda 0} = \textbf{ 0.6209} \pm \textbf{ 0.1654}$. CosmicANNEstimator-SN processed 1048 Type Ia Supernova data points, producing $\Omega_{m0} = 0.2830 \pm 0.0588$ and $\Omega_{\Lambda 0} = 0.9025 \pm 0.1089$. The CosmicANNEstimator-Hubble model demonstrates results consistent with traditional MCMC methods, while the best-fit values predicted by CosmicANNEstimator-SN deviate slightly from their MCMC values. Both models provide substantial computational advantages.

We summarize the  findings and advantages of our method as follows:

\begin{enumerate}

	\item  Unlike MCMC approaches requiring explicit likelihood functions, these neural network-based methods learn parameter relationships directly from training data, providing particular utility for complex, high-dimensional astronomical datasets where likelihood function specification becomes challenging.
	
	\item  The complete training procedure for the 100-member ensemble of CosmicANNEstimator-Hubble required approximately 180 minutes on an AMD64 Family 25 Model 80 CPU system (8 cores, 16 threads, 3.2 GHz processor). Similarly, the CosmicANNEstimator-SN training procedure required approximately 240 minutes on the same computational setup. Following the training phase, parameter estimation for new datasets is achieved with negligible computational overhead for both models, representing a substantial improvement over MCMC methods which typically require significant computational resources and extended runtime for each new analysis.
	Parameter estimation for the Hubble dataset containing 31 observational points required approximately 20 seconds of computation time, while the Supernova dataset comprising 1048 data points processed through CosmicANNEstimator-SN completed parameter inference within approximately 40 seconds.
	
	\item Our neural network approach demonstrates a favorable computational profile, with resource allocation primarily concentrated in two phases: initial training set generation and network training. The efficiency of the ANN approach becomes particularly relevant in the context of modern astronomical surveys, where data volumes continue to expand. While the initial investment in training time is non-negligible, the subsequent rapid parameter estimation capabilities provide a compelling advantage for large-scale cosmological analyses. This computational efficiency positions neural network methodologies as particularly valuable tools for current and future astronomical surveys where rapid analysis of extensive datasets is essential.
	
	\item In this work, we describe our framework as a likelihood-evaluation-free inference method, since the inference stage avoids explicit likelihood evaluations or statistical sampling that are central to traditional Bayesian or MCMC approaches. Instead, the ANN provides parameter estimates and uncertainties through a predictive ML-based mapping learned from simulations generated under the $\Lambda$CDM model. This differs from fully simulation-based inference (SBI) frameworks~\citep{2016arXiv160506376P,2020PNAS..11730055C}, which can treat the forward model as a true black box, whereas our approach remains anchored to analytic cosmological equations for simulation generation. The ML approaches uncover direct mapping between input and targets to learn hidden patterns between them, providing a complementary mechanism to traditional likelihood-based methods for testing the robustness of scientific results.
	
	\item The CosmicANNEstimator models demonstrate excellent predictive performance within training boundaries. Extrapolation beyond these limits yields poor performance, confirming that reliable parameter estimation is restricted to the training parameter space.
	
	\item Our CosmicANNEstimator, like most deep neural networks, functions as a black box where the internal decision-making is not directly interpretable. While MCMC follows explicit likelihood-based reasoning that can be traced, the ANN learns complex non-linear mappings between data and parameters without clear physical interpretation. In this work, we focused on demonstrating numerical consistency with MCMC and computational efficiency, while improving interpretability—through feature attribution methods or physics-informed architectures—remains an important direction for future development
	
	\item Our study identifies several key advantages of ANN-based parameter estimation over traditional MCMC approaches in cosmological contexts. The ability of neural networks to perform parameter inference with minimal computational requirements establishes them as exceptionally suitable tools for large-scale cosmological studies requiring rapid analysis of multiple datasets.
	
	\begin{itemize}
		\item  Modern and upcoming astronomical surveys (such as LSST~\citep{2019ApJ...873..111I}, Euclid~\citep{2025A&A...697A...1E}, and Roman Space Telescope~\citep{2022MNRAS.512.5311W}) will generate unprecedented volumes of data requiring rapid analysis. ANN framework can process thousands of datasets in the time required for a single MCMC analysis, making it ideal for survey forecasts, systematic studies across parameter spaces, and batch processing of observational catalogs.
		
		\item The instantaneous nature of ANN prediction enables real-time cosmological parameter estimation from new observations. This capability is particularly valuable for time-critical applications, adaptive observing strategies, and interactive analysis workflows where immediate feedback is essential.
		
		\item While MCMC requires substantial computational resources for each individual analysis,  ANN approach front-loads the computational investment during training, then provides essentially cost-free inference thereafter. This amortization of computational cost becomes increasingly advantageous as the number of analyses grows, making ANNs particularly cost-effective for large-scale cosmological studies.
	\end{itemize}
	
	To summarize, the advantages of ANN approach, Neural network method is particularly advantageous when: (1)analyzing large volumes of data from modern astronomical surveys; (2)requiring rapid parameter estimation for multiple datasets; (3)working with high-dimensional observational data where likelihood function specification becomes challenging; (4)conducting survey forecasts requiring analysis of thousands of simulated datasets; (5)implementing real-time or interactive analysis workflows; and (6)performing systematic studies across large parameter spaces where computational efficiency is paramount.

\end{enumerate}

A significant limitation of our current approach is the absence of full parameter covariance information. The heteroscedastic loss function provides only diagonal uncertainty estimates and cannot capture parameter correlations such as the well-known $H_0$-$\Omega_{m0}$ degeneracy. This represents an important area for future development, potentially through approaches such as Cholesky decomposition of covariance matrices or normalizing flow-based posterior modeling.

Also, as a future extension of this work, the CosmicANNEstimator framework could be expanded to perform joint analysis of Type Ia Supernova (SN Ia) and Hubble datasets. This enhanced approach would potentially improve the precision of parameter constraints for the $\Lambda$CDM model.


\end{document}